\documentclass[12pt,onecolumn]{IEEEtran} 
\usepackage{natbib}
 \bibpunct[, ]{(}{)}{,}{a}{}{,}%
 \def\newblock{\ }%
\usepackage{graphicx,subfigure}
\usepackage{enumerate,latexsym}
\usepackage{amssymb,amsfonts,amsmath,color}
\usepackage{url}
\usepackage[normalem]{ulem}

\usepackage{natbib}
 \bibpunct[, ]{(}{)}{,}{a}{}{,}%
 \def\newblock{\ }%

\newcommand{\BEQA}{\begin{eqnarray}}
\newcommand{\EEQA}{\end{eqnarray}}
\newcommand{\define}{\stackrel{\triangle}{=}}

\newcounter{cnt:rmn}
\begin{document}
\title{A Game-Theoretic Analysis of Competitive Editing in Wikipedia: Contributors' Effort to Influence Articles and the Community's Attempt to Ensure Neutrality}
\author{Santhanakrishnan Anand, Ofer Arazy, Narayan B.~Mandayam and Oded Nov
\thanks{S.~Anand is with the Department of Electrical and Computer Engineering,
New York Institute of Technology, New York, NY 10023, USA. E-mail: {\tt asanthan@nyit.edu}}
\thanks{O.~Arazy is with the University of Haifa, Israel, E-mail: {\tt oarazy@is.haifa.ac.il}}
\thanks{N.~B.~Mandayam is with the Wireless Information Networks Lab (WINLAB), Rutgers
University, North Brunswick, NJ 08902, USA. E-mail: {\tt narayan@winlab.rutgers.edu}}
\thanks{O. Nov is with the Department of Technology Management, New York University,
Brooklyn, NY 11201, USA. E-mail: {\tt onov@nyu.edu}}
\thanks{This work was supported in part by the US National Academies Keck Futures Initiative (NAKFI)}
}
\maketitle
\begin{abstract}
Peer production, such as the collaborative authoring of Wikipedia articles, 
involves both cooperation and competition between contributors, and we focus 
on the latter. As individuals, contributors compete to align Wikipedia articles 
with their personal perspectives. As a community, they work collectively to 
ensure a neutral point of view (NPOV). We study the interplay between 
individuals' competition and the community's endeavor to ensure neutrality. 
We develop a two-level game-theoretic model, modeling the interactions of 
ownership-motivated individuals and neutrality-seeking communal mechanisms 
as a Stackelberg game. We present our model's predictions regarding the relation 
between contributors' effort (i.e. typical size of edit) and benefits (i.e. the portion 
of the article they eventually ``own''). We validate the model's prediction through 
an empirical analysis, by studying the interactions of 219,811 distinct contributors 
that co-produced 864 Wikipedia articles over a decade. The analysis and empirical 
results suggest that contributors who make large edits (``creators'') eventually lose 
the article's ownership to those who refine the articles and typically make smaller 
edits (``curators''). Whereas neutrality-seeking mechanisms are essential for 
ensuring that ownership is not concentrated within a small number of contributors, 
our findings suggest that the burden of excessive governance may deter contributors 
from participating.
\end{abstract} 
\baselineskip 25pt
\section{Introduction}\label{sec:intro}
Over the past two decades 
online information goods created 
and used by millions of people, such as open source software and Wikipedia, 
have captured the attention of 
researchers in a variety of disciplines. 
Research in the area has studied communities' 
ability to facilitate the creation of high-quality information goods 
and govern large-scale collaboration 
\citep{Benkler2006,Michelucci2016,Giles2015,Gavalda2014,Sauerman2015}. 
Our focus is Wikipedia, one of the most prominent 
examples of peer-production. Wikipedia recruited 
over 23 million volunteers to produce hundreds of millions of encyclopedic 
entries in 287 languages. As Wikipedia has become one of the 
most popular information sources on the web and the destination 
most internet users turn to when they seek information \citep{Halfaker2012}, 
the quality of its articles has been the topic of public debate. 
Wikipedia is based on wiki technology: a web-based collaborative 
authoring tool that allows contributors to add new content, 
append existing content, and delete or overwrite prior 
contributions \citep{LeufCunningham2001}. 
Wikipedia articles evolve through the continuous additions, deletions, 
and shaping existing content (rewriting, reorganizing, and integrating). 
Wikipedia has developed a 
complex organizational structure, where some contributors tend 
to focus on adding new content (i.e., ``builders''), others are more involved in 
the reorganization and ``refactoring''\footnote{``Refactoring'' is a term borrowed from 
software development and entails the restructuring of existing 
computer code: changing the organization without changing its 
external behavior.} of others' contributions (i.e., ``fixers''),
while others take responsibility for 
quality assurance and fighting vandalism 
\citep{Butler2007}. 

Given their reliance on volunteers, peer production projects such 
as Wikipedia need to attract and retain participants with various 
interests and skill-sets.  
One of the key motivating factors is 
contributors' desire to express their knowledge 
and perspective on the topic by shaping articles' contents.
Prior studies have shown that often indirect 
personal benefits underlie such contributions. For example, a contributor 
may be motivated by the desire to increase her own status and reputation 
within the community 
or promote her own 
worldview in published articles \citep{Edwards2013,Richards2014}. 
Consequently, when many contributors participating in a collective authoring effort attempt 
to promote their perspectives, competition between 
viewpoints is inevitable, and is likely to result in biases \citep{GreensteinZhu}.

Whereas most research in the area emphasized the \emph{cooperative} aspects in 
Wikipedia's co-production, answering the call to shift the focus in online 
collaboration research to issues of scarcity and \emph{competition} 
\citep{Wang2013}, the current study focuses on contributors'
competition over the content in Wikipedia articles. Building on the theory of 
psychological ownership \citep{Pierce2001}, we use the terms ``ownership of articles'' 
and the attempt to ``own articles'' to refer to the way in which contributors imbue 
articles with their personal viewpoints. In developing our conceptualization,
we build on prior studies in the area, which demonstrate that competition and 
struggle over the views expressed in articles is a salient feature of Wikipedia's
co-production \citep{Kane2014,Arazy2020,Bidar2020,Young2020}.
These prior studies have discussed individual contributors' attempt to 
influence articles' contents. However, in a large system such as Wikipedia, 
where participants' contributions tend to be in reference to some existing content 
(contributed by others), knowing the effects of contributors' actions on others is 
essential for understanding the system's dynamics. We, thus, extend prior 
research by investigating the effects of collaborators' behavior on others' 
actions and studying the system's dynamics. 

Contributors' dynamics does not happen in a void. Instead, the community 
works to facilitate and govern the collaborative production process. 
Wikipedia's governance mechanisms were investigated in detail, 
e.g. \citep{SchroederWagner2012,aaltonen2015,Auray2012}. Wikipedia 
has extensive mechanisms to ensure that the co-produced content is of high 
quality, including norms, policies, and technical tools for facilitating collaboration, 
resolving conflicts and fighting vandalism \citep{Forte2009}. In particular, Wikipedia 
emphasizes objectivity (Neutral Point of View; NPOV) as a central pillar of the 
community\footnote{\url{http://en.wikipedia.org/wiki/Wikipedia:FivePillars}}  
and lists ``susceptibility to editorial and systemic bias'' as one of the key aspects of its 
quality assurance work\footnote{\url{http://en.wikipedia.org/wiki/ReliabilityofWikipedia}}  
\citep{GreensteinZhu2012,GreensteinZhu,GreensteinZhu2017,GreensteinZhu2018}.
Of particular relevance to our investigation are studies of governance mechanisms 
intended to ensure neutrality and counter attempts to bias Wikipedia articles 
\citep{Hassine2005,Young2020}. To date, these studies have fallen short of linking 
governance mechanisms to individuals' behavior.

 The overarching objective of our study is to investigate the interplay between governance and 
contributors' competition in peer-production (and specifically, in Wikipedia). 
Game theory, ``the study of mathematical models of conflict and cooperation 
between intelligent rational decision-makers'' \citep{Myerson1991}, 
is well suited for studying competitive dynamics and thus may reveal insights into peer production. 
Despite its demonstrated utility in studying cooperation and competition,
the use of game theory 
in peer production research has been rather limited \citep{JainParkes2009}.
In this study, we develop a two-step game-theoretic model that captures both competition 
between contributors over content ownership and the community's governance efforts to 
maintain neutrality. Contributors in this game theoretic analysis balance the costs of 
participating in Wikipedia's collaborative-authoring process (including the costs of 
editing articles, coordinating work, and complying with the community's governance) with the 
benefits (i.e. ``owning'' portions of an article such that it reflects the contributor's viewpoint or 
agenda). We estimate the effort of editing articles through edits' size, and for convenience 
we refer to those who regularly make large-size edits as ``creators'' and to those who 
typically make small-size edits as ``curators''  
\citep{Kim2017,Rainie2012,Hill2017,Hull2013}\footnote{To verify the validity of the 
``creator''/``curator'' labels, we performed 
an analysis on the dataset from \citep{Arazy2010}, calculating each contributor's 
average edit size and profiling the types of activities that she makes. We found that: 
(a) the larger the relative proportion of the ``Add Content'' and ``Structural Changes'' 
edit types in a contributor's profile, the larger is the contributor's average edit size; and 
(b) the larger the proportion of ``Corrections'' and ``Add References'' edit types, 
the smaller is the contributor's average edit size. Together, these patterns suggest 
that contributors with a large edit size are the ``creators'' who structure and article 
and add much of the content, whereas those making small edit size are ``curators'' 
that mostly engage in refining existing content.} (these labels are inconsequential to our  
modeling and empirical analysis). We employ a ``content  ownership'' algorithm from a 
prior study \citep{Arazy2010}, and operationalize the community's goal of ensuring neutrality 
in terms of reducing contributors' concentration - i.e. maximizing entropy of content 
ownership. We report on a computationally-heavy empirical analysis of $219,811$ 
distinct contributors co-producing $864$ articles that provides support for our model's 
predictions. Processing the seven-hundred-thousand revisions for measuring effort and 
contributor's content ownership demanded that we employ a powerful computational infrastructure.

Intuitively, we expect that the ``creators'' who add much content would retain ownership 
of the articles. However, surprisingly, the key results of our game-theoretic analysis are:
\begin{itemize}
\item Analytically, we find that under the governance mechanisms, the fractional content owned 
by a contributor within a focal article is negatively correlated with the average size of her edits. 
In essence, those who regularly make large-size edits (i.e. ``creators'') loose ownership of the 
article to those who typically exert little effort and make small edits (i.e. ``curators''). 
As a consequence, when the competition over ownership unfolds, i.e., 
the number contributors in an article
is significant, only content added by those whose standard edit size is below the 
group's average survives Wikipedia's refactoring process. Empirically, we corroborate this result. 
\item Analytically, we show that governance should be curtailed to a maximum limit, 
beyond which it discourages contributors from making contributions to an article, 
bringing the co-production process to a halt.
\end{itemize}

\section{Related Work}\label{sec:related}
Scholarly accounts of peer-production initiatives such as Wikipedia 
tend to emphasize the collaborative aspects 
\citep{Arazy2011,Benkler2006,KitturKraut2010,KitturSuh2007,RansbothamKane2011}.
However, competition between collaborators cannot be overlooked \citep{Lasfer2019}. 
Given that Wikipedia has become the primary entry point for those seeking information on the 
web \citep{Halfaker2012}, the impact, and thus the associated benefit, of shaping an article's 
contents can be high \citep{das2013}, as demonstrated by commercially sponsored attempts 
to bias Wikipedia's content \citep{Edwards2013,Richards2014}. Recent news stories
 reveal that commercial entities - often through agents, such as PR agencies or political 
 operatives - have been attempting to manipulate Wikipedia articles' content 
 \citep{McKenzie2012,Golumbia2013,Richards2014}. Such manipulations attempt to portray 
 the interested parties favorably, and have been carried out over years, often employing sophisticated methods (e.g. using multiple accounts, building trust in the community and gaining access to special privileges) \citep{Oppong2014}. Wikipedia fights these manipulations through social, technical and legal means \citep{Edwards2013}.

Additionally, competitions over content can also originate from well-intentioned contributors. 
Prior studies have looked at conflicts of opinions between contributors \citep{Arazy2011}, 
focusing on conflict resolution mechanisms and the impact of conflicts on the article's quality. 
More recently, several research
works explicitly discuss contributors' attempt to influence articles. 
For example, \cite{Umarova2019}, 
\cite{Hube2017}  and \cite{Hube2018} 
describe various forms of biases and violations 
of Wikipedia's NPOV policy, and propose automatic methods for detecting these biases. 
Additional very recent studies empirically demonstrated that Wikipedia contributors 
often shape an article in a way that is aligned with their worldview, such the articles 
often evolve through the ``pulling'' of the article towards the direction that represents 
the editor's perspective. \cite{Arazy2020} state that ``contributors to a focal artifact manipulate 
the article according to their particular viewpoints, thus pulling the artifact's trajectory in different directions'' 
(p. 2014) and then suggest that such ``pulls'' may possibly stem from ``$\cdots$ 
a more deliberate attempt to shape the artifact according to the contributor's personal vision.'' 
(p. 2016). Together, these studies demonstrate that competition and struggle over the views 
expressed in articles is a salient feature of Wikipedia's co-production.

Psychological ownership theory \citep{Pierce2001} may provide a potential explanation 
for why compete to own pieces of an article. According to this theoretical framework, 
actors that occupy a shared social space can easily form an ownership feeling over a 
target (in this case, a Wikipedia article) if they invest much time or energy on it, are familiar 
with it, or have control over it. This is particularly pertinent in cases where the individual 
is the originator of the target (or portion of that target), as in the case of contributing 
content to a Wikipedia article. A contributor of content to the article will feel the content 
is her personal psychological property, and subsequently will be unwilling to share the 
target of ownership with others or lose control over it \citep{Pierce2003}.

Territoriality theory \citep{Brown2005} may further explain how psychological 
ownership could result in competition over an article's contents. According to this 
theory, the stronger an individual's psychological ownership of an object, the greater 
the likelihood he or she will treat that object as his or her territory. Territoriality behavior 
is likely to emerge as community members work together to accomplish a common goal. 
It serves a communicative function, by signaling to others one's stake to a territory or 
an object. Consequently, territoriality may be expressed through the defense of one's 
turf from perceived invasions \citep{Brown2005}. From such a viewpoint, if an individual 
experiences a strong feeling of ownership for the content that she has contributed, 
she may exhibit territorial behavior over that section of the article, which subsequently 
leads her to defend the ``owned'' content from others' attempt to change or overwrite it
\citep{Halfaker2009,Kriplean2007,Scott2004,ThomSantelli2009}\footnote{Please note that 
contributors may also try to protect a particular version of the article and exhibit territorial 
behavior even if they did not edit the article, for example in protecting the content of 
like-minded contributors.}. 
\cite{Kane2014} 
report that roughly two-thirds of the collaborative-authoring patterns in the Wikipedia article 
they had investigated entailed ``defensive filtering''. They explain that in the defensive
 filtering pattern ``The focus of production was on protecting the content that has been 
 created by the co-production community'' and that the nature of interaction is defensive 
 and combative. Contributors may become committed to stabilizing knowledge already 
 co-produced. When assessing whether a change to the article fits with the article's 
 current articulation and ensuring that it does not ``violate organizational and content decisions 
 already made'' \citep{Kane2014}.  \cite{Bidar2020} describe power dynamics in 
 Wikipedia whereby certain contributors may attempt to dominate an article, and block alternative 
 viewpoints. \cite{Young2020} describe such a domination dynamic that is related 
 to gender bias, where core contributors wrestle against peripheral contributors to shape 
 articles according to their viewpoints. 

While territoriality may be linked to positive behaviors, such as increased motivation 
and commitment and could help in the organization of work, it raises concerns for 
competitive tensions and for potential for biases in the co-produced knowledge-based 
goods. For example, \cite{Halfaker2009} show that territorial behavior is directly linked 
to turf wars (namely, the revert of a Wikipedia article to its previous version). 
Furthermore, territorial behavior may deter members' participation \citep{ThomSantelli2009}. 

Distributed collaborative communities often employ policy and community 
norms in order to prevent the expressions of such behavior. The Wikipedia 
community explicitly discourages contributors from taking ownership of their 
work\footnote{Wikipedia founder, Jimmy Wales,
speaks openly against the notion of ownership. For example, see a talk in the $21^{st}$ Chaos 
Communication Congress in December 2004 (talk titled ``Wikipedia Sociographics''); \url{http://ccc.de/congress/2004/fahrplan/event/59.en.html}.}, 
and the ``Ownership of Content'' policy\footnote{\url{https://en.wikipedia.org/wiki/Wikipedia:Ownership_of_content}
retrieved February $8$, $2021$}  specifically points out that being 
a primary contributor is not grounds for asserting possession of an article:

\hangindent=1cm
``\emph{All Wikipedia content $-$ articles, categories, templates, and 
other types of pages $-$ is edited collaboratively. \textbf{No one} [emphasis 
in source], no matter how skilled, or how high standing in the community, has the right 
to act as though they are the \textbf{owner} [originally emphasized
in italics] of a particular page. $\cdots$ 
Some contributors feel possessive about material they have contributed to 
Wikipedia. $\cdots$. 
Believing that an article has an owner of this sort is a common mistake 
people make on Wikipedia}''

How are socio-technical governance mechanisms used to mitigate 
the effects of territoriality? Large-scale social production systems, such 
as Wikipedia, require governance mechanisms in order to direct the 
integration of dispersed knowledge resources in the process of value 
creation. Without such mechanisms, coordinating the work of assessing, 
selecting, shaping, integrating and oftentimes rejecting contributors' 
postings would be impossible \citep{aaltonen2015}. Wikipedia employs 
a community-based governance model that is based on egalitarian 
principles, rather than on formal contracts and hierarchies \citep{Arazy2020}. 
Over the years, Wikipedia has developed an extensive set of mechanisms 
to ensure that the co-produced content is of high quality, including norms, 
policies, and technical tools for facilitating collaboration, resolving conflicts 
and fighting vandalism \citep{SchroederWagner2012,Forte2009,Auray2012}. 
What is unique to Wikipedia governance model is that
 it is collective and dynamic. That is, the community organically develops 
 policies and procedures, and these continuously undergo changes by 
 community members \citep{aaltonen2015}. Thus, governance is collective 
 in the sense that it does not depend on any particular individual or a group 
 of individuals within the system.

Central to Wikipedia's governance is the Neutral Point of View 
(NPOV) principle \citep{Hassine2005,GreensteinZhu2012,GreensteinZhu,GreensteinZhu2017,GreensteinZhu2018,Young2020}, 
which promotes 
objectivity and acts to ensure that  no one particular viewpoint dominates an 
article. NPOV is one of the five fundamental principles (or pillars) of Wikipedia. 
The principles\footnote{\url{https://en.wikipedia.org/wiki/Wikipedia:Five_pillars} 
retrieved February 8, 2021.} states that:

\hangindent=1cm ``\emph{We strive for articles that document and explain major points of 
view, giving due weight with respect to their prominence in an impartial tone. 
We avoid advocacy and we characterize information and issues rather than 
debate them. In some areas there may be just one well-recognized point of 
view; in others, we describe multiple points of view, presenting each accurately 
and in context $\cdots$ Editors' personal experiences, interpretations, or 
opinions do not belong.}''

Contributors to Wikipedia are requested to edit articles with a neutral 
perspective, representing views fairly and trying to avoid bias\footnote{Wikipedia 
founder, Jimmy Wales, commented in a 
mailing list posting that ``If a viewpoint is held by an extremely small (or vastly limited) minority, it does 
not belong in Wikipedia regardless of whether it is true or not and regardless 
of whether you can prove it or not.'' (\url{https://lists.wikimedia.org/pipermail/wikien-l/2003-September/006715.html}; retrieved February 8, 2021)}. However, 
biases, the opposite of NPOV, may arise when a subjective information 
is too difficult to verify, when an issue is too complex to be fully represented 
(whereby generating a consensus requires considerable effort and expertise), 
or in cases when contributors intentionally try to impose their own ideology 
in an attempt to influence readers' viewpoints \citep{GreensteinZhu,Arazy2020,Young2020}. 
Enforcing NPOV has become a central concern for 
Wikipedians, and many of the discussions on articles' coordination spaces 
(i.e. `talkpages') \citep{Arazy2011} and dedicated Internet Relay Chat (IRC) 
channels \citep{GreensteinZhu} concern whether particular section of an 
article reflects the NPOV principle.
Editorial activity is often not distributed evenly within a Wikipedia article, 
such that a small share of the contributors makes a substantial share of the 
revisions. Thus, in enforcing neutrality, the community tries to ensure that 
the ``ownership'' of an article's content is not concentrated in the hands 
of few contributors.
\section{Research Question}\label{sec:rq}
The scholarly literature on peer production has not investigated 
the interplay between individuals' competitive dynamics and the 
socio-technical governance mechanisms. 
Our goal in this paper is, thus, to investigate Wikipedians' 
competition over content ownership and to explore the way in which this 
competitive dynamic is shaped by the community's effort to ensure neutrality. 
While we acknowledge that contributors to Wikipedia also exhibit cooperative 
behavior, our focus here is the competitive dynamics. We see two potential 
mechanisms by which NPOV enforcement could affect the competitive 
dynamics. First, the efforts to ensure neutrality could curtail excessive 
one-sided ownership of an article, and consequently directly impact 
contributors' attempts to own article fractions.  Second, adhering to 
norms, policies and procedures may tax contributors, increasing the 
effort associated with participation in co-production. 
We, thus, pose the following two research questions:\\
\hangindent=1cm
\emph{RQ1: How do contributors balance and costs and benefits of peer-production? 
More specifically, what is the relation between the efforts that contributors regularly exert 
when editing an article and their fractional ownership of that article?} \\
\hangindent=1cm
\emph{RQ2: How do governance mechanisms affect the competitive dynamics underlying 
Wikipedia's co-production process? In particular, how does the community's attempt to 
ensure a neutral point of view affect the relationship between contributors' efforts in editing 
articles (i.e. whether they are characterized as ``creators'' or as ``curators'') 
and the amount of content that they ``own'' within a particular article?}
\section{A Game Theoretic Model for Competition and Governance}\label{sec:game}
\begin{figure*}[htbp]
\begin{center}
\includegraphics[width=4.5in]{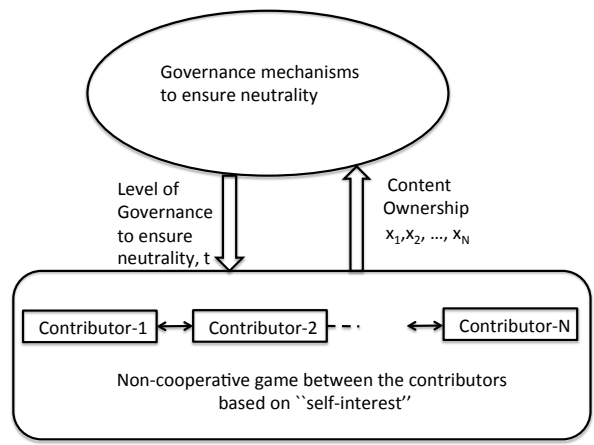}
\caption{\label{fig:basic} A high level schematic 
of the interactive model between Wikipedia's governance to ensure neutrality of an article
and the competition between contributors driven by self-interest to own as much content in an article
as possible.} 
\end{center}
\end{figure*}

A game-theoretic model should capture essential features of the modeled system. 
Whereas we acknowledge that socio-technical systems such as Wikipedia are complex 
and exhibit both cooperative and competitive dynamics, here we attempt to only capture 
Wikipedia's competitive dynamics. Specifically, our game-theoretic model aims at 
representing $(a)$ contributors' competition to ``own'' articles'' contents and $(b)$ the 
community's attempt to ensure that articles present a neutral and balanced perspective. 
Our model assumes that a contributor's goal is to maximize ownership of article sections, 
such that content ``owned'' by that contributor survives multiple rounds of revisions of 
the article. Our model also assumes that the objective of Wikipedia's governance 
mechanisms enacted by the community is to ensure the neutrality of content, 
such that no single contributor takes ownership of large portions of an article 
(please refer to Appendix B for an empirical validation of this assumption).
We acknowledge that Wikipedia's governance has the broader goal of assuring 
content quality and reliability, by ensuring articles' accuracy and completeness. 
Our focus here is on one important dimension of content quality: objectivity or lack of 
bias \citep{Arazy2011}. Beyond the importance of this particular dimension of articles' 
quality (as evident by the NPOV pillar), objectivity is directly impacted by contributors' 
effort to ``own'' content. We model the interactions between the governance mechanisms 
and contributors as a Stackelberg (leader-follower) game \citep{Vonstackelberg1952,Simaan1973}, 
where the community is the leader who determines a level of governance that increases the 
neutrality of an article. Given a set level of governance, contributors' interactions are modeled 
as a non-cooperative game, where each contributor's objective is to maximize her content 
ownership. A high level schematic of the interactive model between the community's goal of 
ensuring neutrality and contributors? competition over content ownership is shown in Figure \ref{fig:basic}.

Peer-production on Wikipedia is a complex socio-technical process and contributions 
occur asynchronously over time. 
Our strategy for capturing these dynamics is temporal bracketing: 
recording a series of yearly ``snapshots'' of the process over a finite time 
horizon \citep{langley99}. We use a static game-theoretic model to study 
these interactions, and test the model against empirical data for each of 
the yearly snapshots. Please note that modeling such a complex process requires some 
simplifying assumptions. We assume that Wikipedia's governance structure,
and in particular the goals of ensuring neutrality, are stable over time (as changes in policies are relatively infrequent). 
As a first-cut approximation, we don't distinguish between the different 
contributors' goals, 
knowledge or roles, as well as assume that no 
learning occurs. It should be stressed that although our model provides an 
approximation to the complexities of peer-production, the model was empirically 
validated with data that captures Wikipedia's actual collaborative-authoring dynamics. 

As mentioned above, our model assumes self-interest driven actions of
contributors, where the goal of individual contributors is to 
advance their personal point of view and ``own'' contents of an article. 
We model the utility of each contributor in the non-cooperative game as her 
fractional ownership in the article\footnote{Using the fractional ownership metric - i.e., the percentage of an article contents 
owned by a particular contributor- we are able to account for differences in length 
between articles. As an example, knowing that one contributed 15\% of an article's 
contents is more informative that knowing that she contributed 8 sentences.
}. Each contributor incurs a cost in making the 
contribution, modeled as the sum total of: (a) the effort expended in producing
content and editing an 
article (this cost element is specific to each contributor); (b) the effort associated with participation 
in coordination and governance work, such as writing policies or sitting on various 
committees (this cost, too, is contributor-specific and is independent of  
her content production activity); and (c) the effort associated with complying 
with governance (similar for all contributors). 
Our objective is then to determine for each contributor the optimal levels of 
activity, so that her net utility is maximized. We model the net utility as the 
difference between the utility (the contributor's fractional ownership) and the 
cost (a measure of all efforts she expended). The optimal strategies are determined 
through the Nash equilibrium of the non-cooperative game that models the interactions 
between contributors. We analyze the conditions for the existence and uniqueness 
of a Nash equilibrium for the game and discuss its implications.

We model the interactions between the governance mechanism - namely, enforcing 
a neutral point of view- and contributors as a Stackelberg (leader-follower) game. 
The role of the neutrality mechanism is to choose the optimal level of enforcement, 
$t$, defined in our model in terms of the effort imposed on contributors to make a 
single unit of contribution (i.e. edit an article). The response of the $N$ contributors 
in the particular article is to make unilateral contributions that yield content ownership, 
$x_1$, $x_2$, $\cdots$, $x_N$, for them, respectively. 
Please refer to Table \ref{tab:notations} 
for the list of notations that are used in our analysis.

We report on an empirical analysis of $219,811$ distinct contributors co-producing 
$864$ articles (lifespan ranges from $129$ to $4078$ days; mean $= 2681$ days) 
that validates our model's predictions. 
We perform the empirical validation for the entire period covered by our data set 
(2001-2012), as well as for each yearly interval independently. The key results of 
our game-theoretic analysis validated by data, include:
\begin{itemize}
\item The neutrality mechanisms that act to avoid uneven distribution in content 
ownership have to be constrained, such that a contributor's effort in complying with 
governance would not exceed a threshold.  
\item Under the neutrality mechanisms, the fractional content owned by a contributor 
is negatively correlated with the effort she expends. 
In particular, the ``creators'' that add large pieces of content loose ownership of 
the article to the ``curators'' who typically engage in small adjustments that require 
little effort. As a consequence, when the competition over ownership unfolds, 
i.e., the number contributors in an article is significant, only content added by those whose standard edit 
requires effort below the group's average (i.e. ``curators'') survives the ongoing 
refinement and refactoring.
\end{itemize}

\subsection{Co-Production: Contributors' Competition}
\label{subsec:ncgame}
We model the interactions of the $N$ content contributors\footnote{Please note that while 
our model assumes a given number of contributors, it does allow for contributors to make 
a choice regarding their participation. Namely, by deciding to own zero content contributors 
are actually deciding not to join the game.} to a particular Wikipedia article 
as a non-cooperative game. The strategy set is 
the content added to the article by each contributor  $x_1$, $x_2$, $\cdots$, $x_N$. 
Contributors' actions are
aimed at maximizing their content owned. Specifically, the actions of contributors 
are changes to the co-created artifact (i.e. ``edits''), whereby each contributor
``owns'' the content she added, and indirectly reducing others' fractional ownership.

The $i^{th}$ contributor has
fractional content ownership,
\begin{eqnarray}\label{eqn:utilityi}
c_i= \frac{x_i}{\sum_{j=1}^N x_j}.
\end{eqnarray}
Contributors' editing actions incur a cost (i.e. the effort 
in explicating knowledge and modifying an article).
Contributors are characterized by their activity profile within Wikipedia. 
In particular, the $i^{th}$ contributor is characterized by the average size 
of his edits to Wikipedia articles, $\beta_i$. Building on work in the area of 
software development which approximated coding effort based on the quantity 
of code produced (i.e. lines of code) \citep{Shihab2013,Mendes2006}, we 
assume that effort is a linear function of edit size, such that the contributor's 
average effort per content contribution is $L\beta_i$. Hence, the effort contributor 
$i$ makes to the focal article is a function of his regular effort expenditure per 
edit, a constant, and the amount of content he contributed to the focal article
i.e., the effort invested in contributing content to an article is $L\beta_ix_i$.

In addition, the 
$i^{th}$ contributor needs to expend additional effort
to overcome the attempt by 
Wikipedia's governance to  ensure neutrality of an article 
(\emph{e.g.}, in learning 
and complying with Wikipedia's rules and policies), 
and yet maintain every unit of her ownership. Therefore,
the $i^{th}$ contributor owning $x_i$ amount of content
also expends an 
effort $tx_i$ to overcome the level of neutrality enforcement, $t$,
(in addition to the effort in editing the article, $\beta_ix_i$).  
A contributor also expends efforts and incurs a cost by participating in 
coordination and administrative work. This cost
element is user specific but fixed for a user (i.e., independent of
the size of the edits they make) and 
we model this fixed cost as $f_i$, for
user $i$.  
The net utility 
experienced by the $i^{th}$ contributor, $u_i$, can be written 
as the difference between utility of contributor $i$, given by 
Equation (\ref{eqn:utilityi}) and the total effort expended by 
contributor $i$. Formally stated: 
\begin{eqnarray}\label{eqn:netutilityi}
u_i=c_i-(L\beta_i+t)x_i-f_i=
\frac{x_i}{\sum_{j=1}^N x_j}
-(L\beta_i+t)x_i-f_i.
\end{eqnarray}
Note that in general,  a convex function of $x_i$ would be the most
comprehensive model for the cost incurred by each contributor, to own $x_i$
amount of content. We consider a linear cost model, which is a special case
of a convex cost function \citep{Luenberger1984}, which we use in Equation (\ref{eqn:netutilityi}).

Each contributor
then determines her optimal amount of contribution, $x_i^*$,
that maximizes, her net utility, $u_i$ \emph{when the contributions
made by all other contributors, $x_j$, $j\neq i$ are fixed.}
It is therefore observed that the net utility obtained by the  $i^{th}$ 
contributor not only depends on the strategy of the $i^{th}$ 
contributor ({\it i.e.}, $x_i$), but also on the strategies of 
all the other contributors ({\it i.e.}, $x_j$, $j\neq i$)\footnote{Therefore, 
$u_i$ is also sometimes denoted as
$u_i(x_i,\mathbf{x}_{-i})$ where $\mathbf{x}_{-i}$ represents
the vector, $\left [ \begin{array}{ccccccc} 
x_1 & x_2 & \cdots & x_{i-1}&x_{i+1} &\cdots & x_N
\end{array}\right ]^T$, i.e., all elements of the vector, $\mathbf{x}
=\left [ \begin{array}{cccc} 
x_1 & x_2 & \cdots & x_N
\end{array}\right ]^T$
except the $i^{th}$ element.}. 
This results in the non-cooperative game of complete information  
\citep{VonNeumannMorgenstern1944} between the contributors. The 
optimal $x_i$, $\forall$ $i$ (denoted as $x_i^*$), which is 
determined by maximizing $u_i$ in Equation (\ref{eqn:netutilityi})\footnote{when 
all other $x_j$'s, $j\neq i$ are fixed, i.e., the vector, $\mathbf{x}_{-i}$ is fixed.}, 
is then the Nash equilibrium of the non-cooperative game where no 
contributor can make a unilateral change. 

The unique Nash equilibrium,
$\mathbf{x}^*=
\left [
\begin{array}{cccc}
x_1^* & x_2^* & \cdots & x_N^*
\end{array}
\right ]^T$, 
can be obtained in closed-form as follows.
$\mathbf{x}^*$ is obtained by determining the best response
for contributor $i$, $x_i^*$ which maximizes its net utility, $u_i$ (given by
Equation (\ref{eqn:netutilityi})) when the contributions made by all other
contributors, i.e., $x_j$, $j\neq i$ are known. However, it is noted that \emph{every
contributor attempts to do the same.} In other words, the unique Nash equilibrium, 
$\mathbf{x}^*=
\left [
\begin{array}{cccc}
x_1^* & x_2^* & \cdots & x_N^*
\end{array}
\right ]^T$,
is the vector that maximizes the net utility, $u_i$, in Equation 
(\ref{eqn:netutilityi}), \underline{$\forall$ $i$}. 
Applying the first order 
necessary condition to Equation (\ref{eqn:netutilityi}), $x_i^*$
is obtained as the solution to
\begin{eqnarray}\label{eqn:firstderivative}
\begin{array}{cc}
\left .
\frac{\partial u_i}{\partial x_i}
\right |_{x_i=x_i^*}=
\frac{\sum_{k = 1\atop k \neq i}^{N} x_k^*}
{\left (\sum_{j = 1}^{N} x_j^*\right )^2} - (L\beta_i+t)=0,
& \forall i
\end{array}
\end{eqnarray}
subject to the constraints
$x_i^*\geq 0$, $\forall$ $i$.

From Equation (\ref{eqn:firstderivative}), we obtain
$\frac{\partial^2 u_i}{\partial x_i^2} = - \frac{2
\sum_{k = 1\atop k \neq i}^{N} x_k}
{\left (\sum_{j = 1}^{N} x_j\right )^3} < 0$,
$\forall$ $i$, when $x_i\geq 0$.
Thus, $u_i$ is a concave function of $x_i$
and $x_i^*$, which solves Equation (\ref{eqn:firstderivative})
subject to $x_i^* \geq 0$, $\forall$ $i$,
is a local as well as a global maximum point.
In other words, 
\emph{the non-cooperative game has a unique Nash equilibrium},
The vector of contributions,
$\mathbf{x}^*=
\left [
\begin{array}{cccc}
x_1^* & x_2^* & \cdots & x_N^*
\end{array}
\right ]^T$, can be
obtained by numerically solving the system of $N$ 
non-linear equations specified by Equation (\ref{eqn:firstderivative}).
However, to study the effect of the contributors' type (i.e. their $\beta_i$'s) 
and the level of neutrality enforcement, $t$, on contributors' strategies, it is 
desirable to obtain an expression that relates the vectors, 
$\mathbf{x}^*$, $\mathbf{x}=\left [x_i\right ]_{1\leq i\leq N}$, 
$\mbox{\boldmath $\beta$}=\left [\beta_i\right ]_{1\leq i\leq N}$ 
and $t$. 

Solving Equation (\ref{eqn:firstderivative}),
the unique Nash equilibrium of the non-cooperative game
can be obtained as\footnote{See Appendix \ref{app:NashEq} for details.} 
\begin{eqnarray}\label{eqn:bistar}
x_i^*=\frac{\sum_{j=1}^NL\beta_j-(N-1)L\beta_i+t}
{\left (\sum_{j=1}^N(L\beta_j+t)\right )^2}.
\end{eqnarray}
Note that the unique Nash equilibrium $\mathbf{x}^*$, is feasible, 
{\it i.e.}, $x_i^* >0$, $\forall$ $i$ if and only if
\begin{eqnarray}\label{eqn:conditionnash}
(N-1)L\beta_i< \sum_{j=1}^NL\beta_j+t\mbox{ or } (N-2)L\beta_i<\sum_{j=1\atop j\neq i}^N L\beta_j +t.
\end{eqnarray}
Intuitively, Equation (\ref{eqn:conditionnash}) implies the following. Let all the other contributors as well as
the neutrality enforcement represent ``adversaries'' of contributor $i$. The right hand side shows the
total adversarial activity level against contributor $i$. The left hand side shows a term similar to
the total  activity level in the favor of contributor $i$ if contributor $i$ were to
duplicate herself against each individual adversary.  Equation (\ref{eqn:conditionnash})
then says that a contributor makes a positive contribution to an article 
only if the contributor has to expend effort in his/her favor, that is
less that the effort or activity level of the adversaries.

The fractional content ownership of contributor $i$ at the Nash equilibrium, 
$c_i^*$, can then be obtained from Equation (\ref{eqn:utilityi}) and 
Equation (\ref{eqn:bistar}) as 
\begin{eqnarray}\label{eqn:uistar}
c_i^*=\left ( \frac{x_i^*}{\sum_{j=1}^N x_j^*}\right )^+=\left [1-\left ( \frac{(N-1)(L\beta_i+t)}{\sum_{j=1}^N(L\beta_j+t)}\right )\right ]^+,
\end{eqnarray}
where for any, $\theta$, $\theta^+=\max(\theta,0)$\footnote{In other words, 
if the term $\theta$ is positive or zero, then use it as it is. If $\theta$ is negative, then
take it to be $0$.}. 
It is observed that the ownership $c_i^*$ is non-zero only 
if Equation (\ref{eqn:conditionnash}) is satisfied, i.e., if the 
Nash equilibrium is feasible.

Note that the
optimal $x_i^*$ in Equation (\ref{eqn:bistar}) and the optimal fractional
ownership, $c_i^*$ in Equation (\ref{eqn:uistar}), do not depend on the
fixed cost, $f_i$. This is because, the fixed cost does not change with
changing $x_i$ (from Equation (\ref{eqn:netutilityi})) and hence has no
bearing on the optimal $x_i^*$. Similarly, from Equation (\ref{eqn:utilityi}),
the fractional ownership, $c_i$, is independent of $f_i$ and therefore,
$c_i^*$ in Equation (\ref{eqn:uistar}) is also independent of $f_i$.
The condition in Equation 
(\ref{eqn:conditionnash}) and the expression in Equation 
(\ref{eqn:uistar}) have the following interesting implications.
\begin{itemize}\itemsep -2pt
\item From Equation (\ref{eqn:uistar}),
the ownership of contributors depend on
the $\beta_j$ of {\it all the contributors}. 
This is intuitively correct in a peer production project 
like Wikipedia because changes to the produced artifact are 
made by multiple contributors and the ownership held by a 
contributor will depend on the activity tendencies of all the contributors 
co-producing the artifact.
\item The expression in Equation (\ref{eqn:uistar}) indicates 
that contributors who tend to make small changes to articles (i.e. ``curators'') are left with
larger fractional
ownership and those who regularly make large edits (i.e. ``creators'') remain with smaller 
fractional ownership, i.e., 
the fractional content ownership is a decreasing function of the 
typical size of one's edits. 
\item In Equation (\ref{eqn:bistar}), the level of 
neutrality enforcement, $t$, appears as a linear factor in the numerator and as a second order
quadratic factor (power of $2$) in the denominator. Therefore, when neutrality enforcement, $t$ increases,
the overall content owned by a contributor decreases. This implies that
when the ``tax'' imposed on contributors in terms of complying with NPOV
norms, policies and procedures is too high it outweighs the benefits 
associated with content ownership, such that contributors stop 
competing for ownership (and in effect, co-production is stalled).
\item Equation (\ref{eqn:uistar}) can be re-written as
\begin{eqnarray}\label{eqn;cistarrewritten}
c_i^*=\left [1-\left ( \frac{L\beta_i+t}{\frac{1}{N-1}\sum_{j=1}^N L\beta_j +\frac{N}{N-1}t}\right )\right ]^+\nonumber\\
=\left [1-\left ( \frac{L\beta_i+t}{\frac{N}{N-1}\frac{1}{N}\sum_{j=1}^N L\beta_j +\frac{N}{N-1}t}\right )\right ]^+.
\end{eqnarray}
Asymptotically, i.e., as the number of contributors, $N$, 
becomes large $c_i^*$ is evaluated by taking  $\lim_{N\rightarrow\infty}$ in 
Equation (\ref{eqn;cistarrewritten}), to obtain
\begin{eqnarray}\label{eqn:largeN}
c_i^*= \left [1-\left ( \frac{L\beta_i+t}{LE[\mbox{\boldmath $\beta$}] +t}\right )\right ]^+\nonumber\\
= \left (\frac{LE[\mbox{\boldmath $\beta$}] -L\beta_i}{LE[\mbox{\boldmath $\beta$}] +t}\right )^+,
\end{eqnarray} 
where $E[\mbox{\boldmath $\beta$}]\define\frac{1}{N}\sum_{j=1}^N\beta_j$,
is the {\it average size of edits} of all contributors making changes to the 
focal Wikipedia article. 
It is observed from  Equation (\ref{eqn:largeN}) that $c_i^*>0$ for only those contributors 
for whom $\beta_i<E[\mbox{\boldmath $\beta$}]$. 
Therefore, only those contributors whose edit size (and thus, edit cost) in making a unit contribution 
is below the average edit cost by contributors working on the focal article, end up with non-zero ownership, 
when the number contributors in an article is significant.
\item The $c_i^*$ given by Equations (\ref{eqn:uistar})  and (\ref{eqn;cistarrewritten}) are 
the exact expressions for the content ownership. 
The asymptotic behavior in Equation (\ref{eqn:largeN}) is used only to reach the conclusion that 
when the number contributors in an article is significant,
only those contributors whose edit sizes are less than the average edit size, survive the competition 
and end up owning non-zero percentage of content in the article. For all the numerical analysis, only Equation (\ref{eqn:uistar}) is used
\end{itemize}

\indent While the above result seems counter-intuitive, it is in fact a result 
of the governance mechanisms and its desire to increase neutrality. 
One might expect that the creators who contribute more content (and in 
the process exert more effort) 
would end up owning much of an article's contents. In contrast, the 
results of our game-theoretic analysis implies that when competing over 
content ownership in the presence of Wikipedia's governance
to ensure neutrality, those contributors making 
on average large contributions (i.e. creators) would eventually not own any content.
Please refer to Appendix \ref{app:notations} for a summary of the notations we use and to 
Appendix \ref{app:NashEq} for details of the Nash equilibrium of the non-cooperative game between contributors.
\subsection{Neutrality Enforcement and its Effect on Co-Production}\label{sec:gov}
The community governance factor 
in our model is the result of contributors' 
aggregate governance effort.
The level of governance, $t$, 
should be chosen so as to maximize the 
effectiveness of peer production. The main objective of
Wikipedia's governance is to reduce bias and ensure that all points of view are reflected in an article.
Wikipedia
policies define Neutral Point of View as a central pillar of 
the community and its quality assurance procedures pay particular 
attention to the elimination of biases. 
Prior studies in the area have 
stressed the importance of the Neutral Point of View norm in Wikipedia's governance 
\citep{Forte2009,GreensteinZhu}. 
Biases occur 
when a single viewpoint dominates the deliberation, or in the case 
of Wikipedia, the article gives more weight to one point of view 
over another. Often, such biases are the result of one contributor 
taking substantial ownership of an article; in contrast, when multiple 
contributors own sections of the article, a more neutral point of 
view is expected. 
Thus, our notion of article neutrality reflects equal ownership of an 
article's contents (and we are less interested here in the particular 
biases that may be present in the content added by each contributor).

Building on information theory \citep{Shannon1948}, 
we use the metric of entropy to estimate the amount of bias in the distribution of content ownership. 
We note that the entropy metric was employed in prior studies in organization science \citep{AnaconaCaldwell1992}, 
and in particular in estimating the distribution of work in Wikipedia \citep{Arazy2011}. 
In information theoretical terms, if a source emits $N$ possible symbols,
$\mathbf{s}=\left [
\begin{array}{ccccc}
W_1 &Ws_2 & W_3 & \cdots & W_N
\end{array}
\right ]^T$ with probability distribution, $p_i$, i.e.,
the source emits symbol, $W_i$, with probability, $p_i\geq 0$, $\sum_{i=1}^Np_i=1$, then
the entropy of the source, $H$
is defined as
\citep{CoverThomas}
\BEQA\label{eqn:entropydefinition}
H=\sum_{i=1}^N p_i\mbox{ln }\frac{1}{p_i}=-\sum_{i=1}^N p_i\mbox{ln }p_i.
\EEQA
In any Wikipedia article, the fractional
ownership of each contributor, $c_i^*$ are all non-negative
and satisfy the condition,
$\sum_{i=1}^N c_i^*=1$. Therefore, page can be viewed as
as information source, the $N$ contributors play the role of the
$N$ possible symbols and the fractional content ownership of
each contributor, $c_i^*$, plays the role of $p_i$. The entropy of the page can then
be written as
\BEQA\label{eqn:pageentropy}
H(t)=\sum_{i=1}^Nc_i^*\mbox{ln }\frac{1}{c_i^*}=-\sum_{i=1}^Nc_i^*\mbox{ln }c_i^*.
\EEQA
In Equation (\ref{eqn:pageentropy}), the entropy, $H(t)$, may depend on 
other factors, e.g., the optimal contributions, $x_i^*$ $1\leq i\leq N$, obtained
from Equation (\ref{eqn:bistar}), which in turn, depend on $\mbox{\boldmath{$\beta$}}$.
However, we represent the entropy as $H(t)$ because the neutrality enforcement, $t$, 
is the the only parameter in $H$ that represents the action taken by Wikipedia's governance
mechanisms. Our model's optimization function for neutrality enforcement 
(maximizing entropy in contributors' ownership) is normalized (each article's entropy 
is calculated in relation to the maximum entropy of any article, which is $1$).

It was shown in information theory \citep{CoverThomas} that the expression
of entropy, $H(t)$, in Equation (\ref{eqn:pageentropy}) is maximized  when
$c_i^*=\frac{1}{N}$, $\forall$ $i$. i.e., when all contributors make equal contribution,
which corresponds to the scenario when there is complete neutrality. Therefore, we use
the entropy in Equation (\ref{eqn:pageentropy}) as the objective function for
determining the optimal level of neutrality enforcement for the Stackelberg game.
Namely, the objective function for neutrality enforcement in our model is to minimize the bias or 
maximize the entropy in content ownership within an article\footnote{In
Appendix \ref{app:JMISDataSet}, we show empirical evidence that there is strong positive
correlation between the entropy of an article and its quality (and hence, its neutrality), thus justifying the . 
objective function in Equation (\ref{eqn:optimalt}).}.
The optimal level of neutrality enforcement, $t$, is the value of $z$ that maximizes
$H(z)$ in Equation (\ref{eqn:pageentropy}), which is obtained by replacing $t$
by $z$ in Equation (\ref{eqn:uistar}). 
Then, 
\BEQA\label{eqn:optimalt}
t=\mbox{arg}\max_z H(z),\mbox{ i.e., }\left . \frac{\partial H}{\partial z}\right |_{z=t}=0.
\EEQA

\begin{figure*}[htbp]
\begin{center}
\includegraphics[width=4.5in]{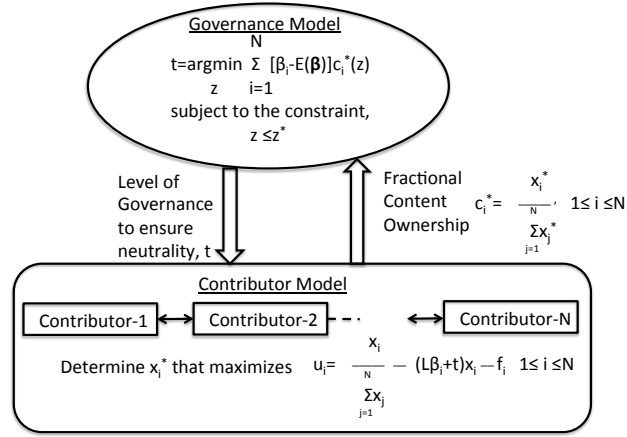}
\vspace{-0.5in}
\caption{\label{fig:actual} The comprehensive Stackelberg game depicting
the complete interaction between Wikipedia's governance to ensure neutrality and 
competition between contributors to maximize content ownership.} 
\end{center}
\end{figure*}

From Equations (\ref{eqn:uistar}), (\ref{eqn:pageentropy}) and (\ref{eqn:optimalt}),
the optimal value of $t$ that maximizes the entropy is the value of $t$ that makes $c_i^*=0$,
$\forall$ $i$, i.e., $t\rightarrow\infty$\footnote{See Appendix \ref{app:Optimalt} for details.}. 
Our analysis indicates that 
while neutrality enforcement levels remain moderate, our prior results from the 
non-cooperative game remain unchanged, such that those making 
above-average contributions end up not owning content. However, 
when neutrality enforcement levels exceed a certain threshold the 
non-cooperative model collapses, ownership becomes a direct function 
of neutrality enforcement (rather than contributors' efforts). Namely, 
from Equation (\ref{eqn:uistar}), when $t>> N E(\mbox{\boldmath{$\beta$}})$
(i.e. when the amount of effort a contributor has to exert to overcome neutrality enforcement is 
significantly- typically one order of magnitude- larger than the aggregate effort 
of all contributors exert in making unit contributions to the article), 
$c_i^*\approx \frac{1}{t}\rightarrow 0$. In other words, ALL 
contributors end up owning an insignificant but equal portion of the article's contents. 
Therefore, 
the optimal value of $t$ is obtained by solving
\BEQA\label{eqn:optimaltwithconstraint}
t=\mbox{arg}\max_z H(z),\mbox{ i.e., }\left . \frac{\partial H}{\partial z}\right |_{z=t}=0,\nonumber\\
\mbox{subject to the constraint}\nonumber\\
t\leq z^*,
\EEQA
where $z^*$ represents the maximum level of neutrality enforcement allowed
in order to maintain the neutrality of an article above a required threshold. 
The resulting Stackelberg game model is depicted in 
Figure \ref{fig:actual}.

The complete model for the principal agent (leader-follower) game between the 
governance mechanism and the competitive self-interested contributors suggests 
that at the Stackelberg equilibrium, governance mechanism operates in such
a manner to result in an implicit outcome that the entropy is kept high, i.e., the
difference between maximum 
and minimum fractional content ownership is kept low, therefore ensuring article's objectivity. 
Further, our model limits governance levels to an upper threshold.
\section{Verification of the Analysis with Empirical Data}\label{sec:results}
To verify the game theoretic analysis, we compared its predictions against 
data from Wikipedia. 
In line with recent studies in the area \citep{Arazy2016}, 
we employed a double-stratified sampling procedure, 
randomly selecting 1000 articles from the January 2012 dump of the English 
Wikipedia. Our strata were based on: $(I)$ the number of revisions
the article has gone throught (i.e., article maturity) 
and $(II)$ the article's topical domains. 
This sampling strategy is important given that collaboration patterns could differ across 
articles in different phases of their life cycle 
across topical domains. 
Given the power 
law distributions in the number of articles' revision \citep{Ortega2008}, we 
used the following four maturity strata: $(a)$ 1-10 revisions; $(b)$ 11-100; 
$(c)$ 101-1000; and $(d)$ more than 1000 revisions. 
The topical strata were 
based on Wikipedia's categorization system, using the main topics 
scheme\footnote{The English Wikipedia main topic categorization scheme 
is developed by the community and is subject to frequent changes; see 
\url{http://en.wikipedia.org/wiki/Category:Main_topic_classifications}.}.
With 4 maturity strata and 25 topical categories, we have 100 cells with 10 
randomly selected articles in each (i.e. 250 articles in each maturity 
stratum and 40 articles in each topical category). Altogether, this 
sample contained: 1000 articles and 721,806 editing activities 
(i.e. article revisions), made by 222,119 contributors\footnote{In line 
with prior works in the area \citep{Arazy2011}, we assume that for 
non-registered contributors, a unique IP address corresponds to a 
distinct contributor. About half of the contributors in our sample are non-registered.}. 
After excluding 
articles with a single contributor, 
we were left with 864 multi-contributor articles, which make  up the 
sample for our analysis. 

We measured contributors' benefit (i.e. ownership of articles' contents) 
and costs (namely, the effort in making an edit) by calculating the fractional 
ownership for each contributor in each article of our sample using the algorithm in 
\citep{Arazy2010}. The content ownership algorithm tracks the evolution of content, 
recording the number of sentences owned by each contributor at each revision, 
until the study's end date. The algorithm uses a sentence as the unit of analysis, 
where each full sentence is initially owned by the contributor who first added it; 
as content on a wiki page evolves, a contributor loses a sentence when more than 50\% 
of that sentence's content (i.e. the words in the original sentence) is deleted or revised by 
others (thus, a contributor making a major revision to an existing sentence can take 
ownership of that sentence). If no single contributor
``owns'' more than 50\% of a sentence, that sentence becomes ownerless. 
The output of the algorithm is, for each contributor, the number of sentences 
(and the fraction of sentences within the focal Wikipedia article) 
originated by her that persisted in the most recent version of the article.

We defined $s_i$ as the number of sentences owned by contributor $i$. 
A contributor's fractional ownership in a particular article is obtained as 
$\frac{s_i}{\sum_{j=1}^N s_j}$.  
Next, we calculated the effort expended by the contributor. 
A contributor's average edit size, $\beta_i$, 
was calculated based on his activity profile across all 
articles within the sample to which he contributed.
The size of edits was calculated using 
the Levenshtein distance \citep{Levenshtein1966}. 
The Levenshtein distance is a string metric used in information theory and 
computer science for measuring the difference between two sequences (or 
between two text segments) and is defined as the minimum number of 
single-character edits (i.e. insertions, deletions or substitutions) required 
to change one word into the other \citep{RistadYianilos1998}. This metric 
has been commonly used in 
the study of Wikipedia to estimate the scope of editing activities \citep{Keith2007}. 
Based on this data, we were able to calculate the following parameters for 
each of the 864 articles in our sample:
\begin{itemize}
\item The number of distinct contributors ($N$)
\item The total number of edits made by the $i^{th}$ contributor ($1\leq i\leq N$), $\zeta_i$, across all of the articles in our sample
\item The size of all edits made by the $i^{th}$ contributor ($1\leq i\leq N$), $e_i$, to all articles in our sample
(summing up the Levenshtein distance 
of all edits made by that contributor).
\end{itemize}	
Using these set of parameters, we compute the average edit size of 
contributor $i$ for unit contribution, $\beta_i$, $\beta_i=\frac{e_i}{\zeta_i}$.
Note that $\beta_i$ is a trait of the contributor, such that it remains the same
for every article that we analyze. 

The number of contributors in the $864-$article sample  was found
to follow a uniform distribution with average number of contributors being
$124.6$ and a standard deviation, $71.9$. We also found that contributors 
are more likely to make smaller edits than larger edits, with the distribution of
$\beta_i$  resembling an exponential probability density function. The average $\beta_i$ 
was 8.4 (measured in Levenshtein distance) and the standard deviation was 2.9. 
The average content ownership in an article for users, $c_i^*$, was found
to have a mean of $8$\% and a standard deviation of $0.55$.

Employing the $\beta_i$'s thus obtained, we use the expression in 
Equation (\ref{eqn:uistar}) to determine our model's prediction for a contributor's 
fractional ownership in a particular Wikipedia page. Equation (\ref{eqn:uistar}) provides us with a 
contributor's expected fractional ownership. We repeated this analysis for all 
Wikipedia articles in our sample and for all contributors in each article. 

We empirically tested our model's predictions regarding the relation between 
contributors' average edit size (i.e. a proxy for the effort typically exerted by a 
contributor when making an edit) and their resulting content ownership. 
and their resulting content ownership. 
In trying to get a better understanding of the temporal dynamics underlying 
the co-production competition over an article's content, we applied a temporal 
bracketing technique by breaking each article into yearly brackets (i.e. 
Year 1 since inception, Year 2, etc.). We then created 
subsets by yearly periods, capturing the states of all articles after 
their first year of operation, second year etc., and calculated 
contributors effort and fractional ownership at the end of each year in an 
article's life. Next, for each of these yearly subsets, we repeated the 
analysis described above, comparing the model's prediction to actual 
fractional ownership. 
\begin{figure*}
\begin{center}
\subfigure[Entire Period of Study]{
\includegraphics[width=3.1in]{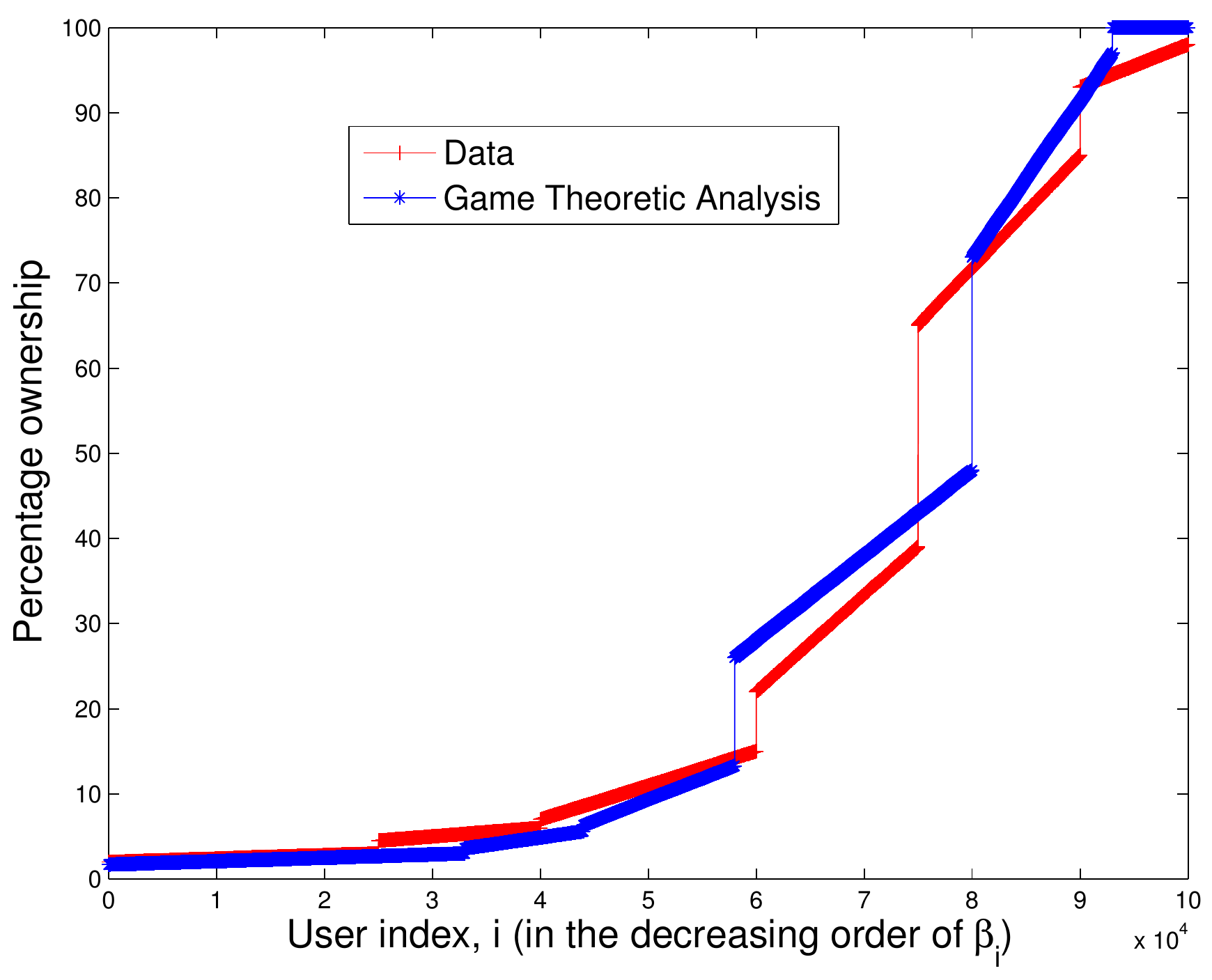}
\label{fig:compare}}
\subfigure[First Year]{
\includegraphics[width=3.1in]{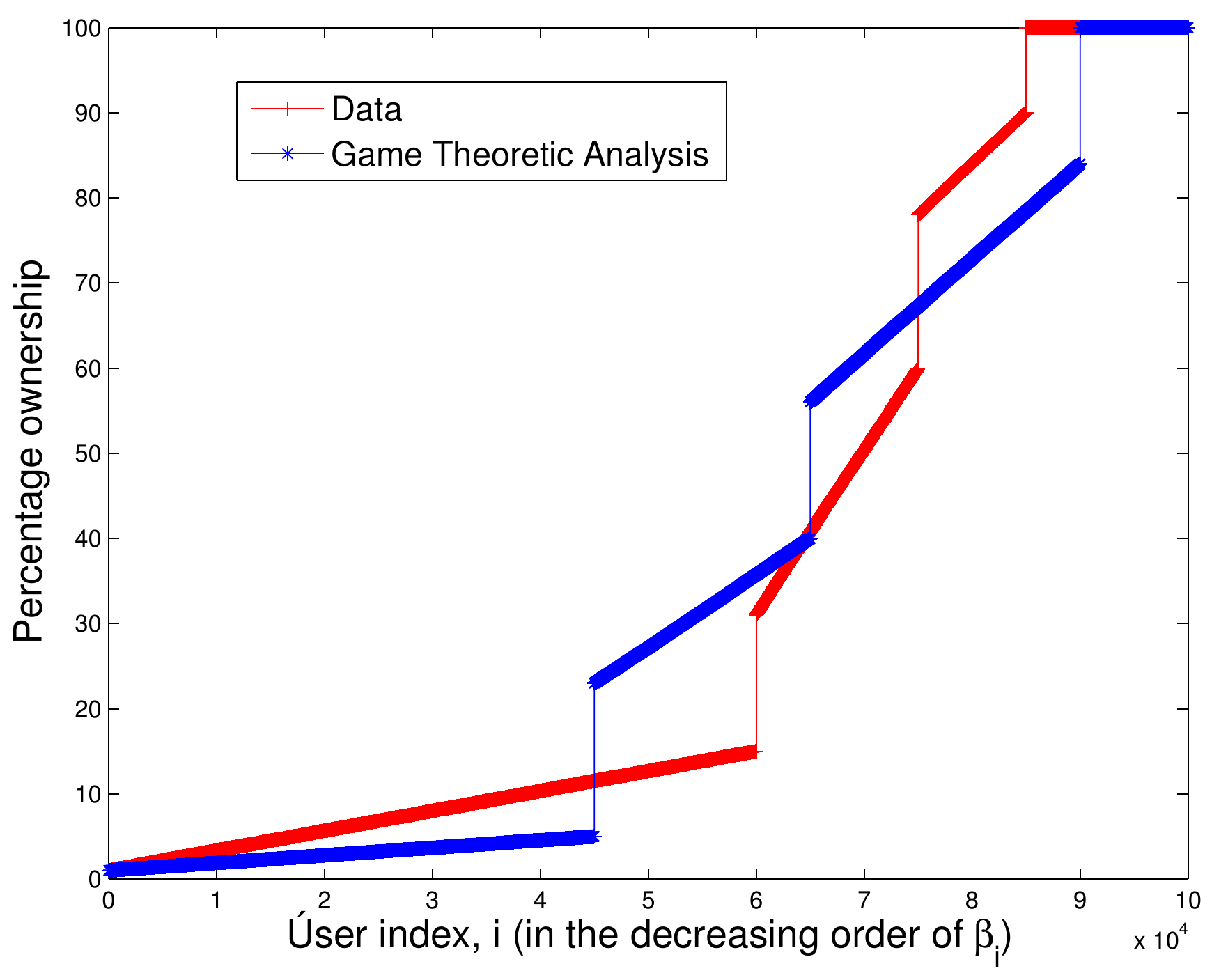}
\label{fig:comparefirst}}
\subfigure[Second Year]{
\includegraphics[width=3.1in]{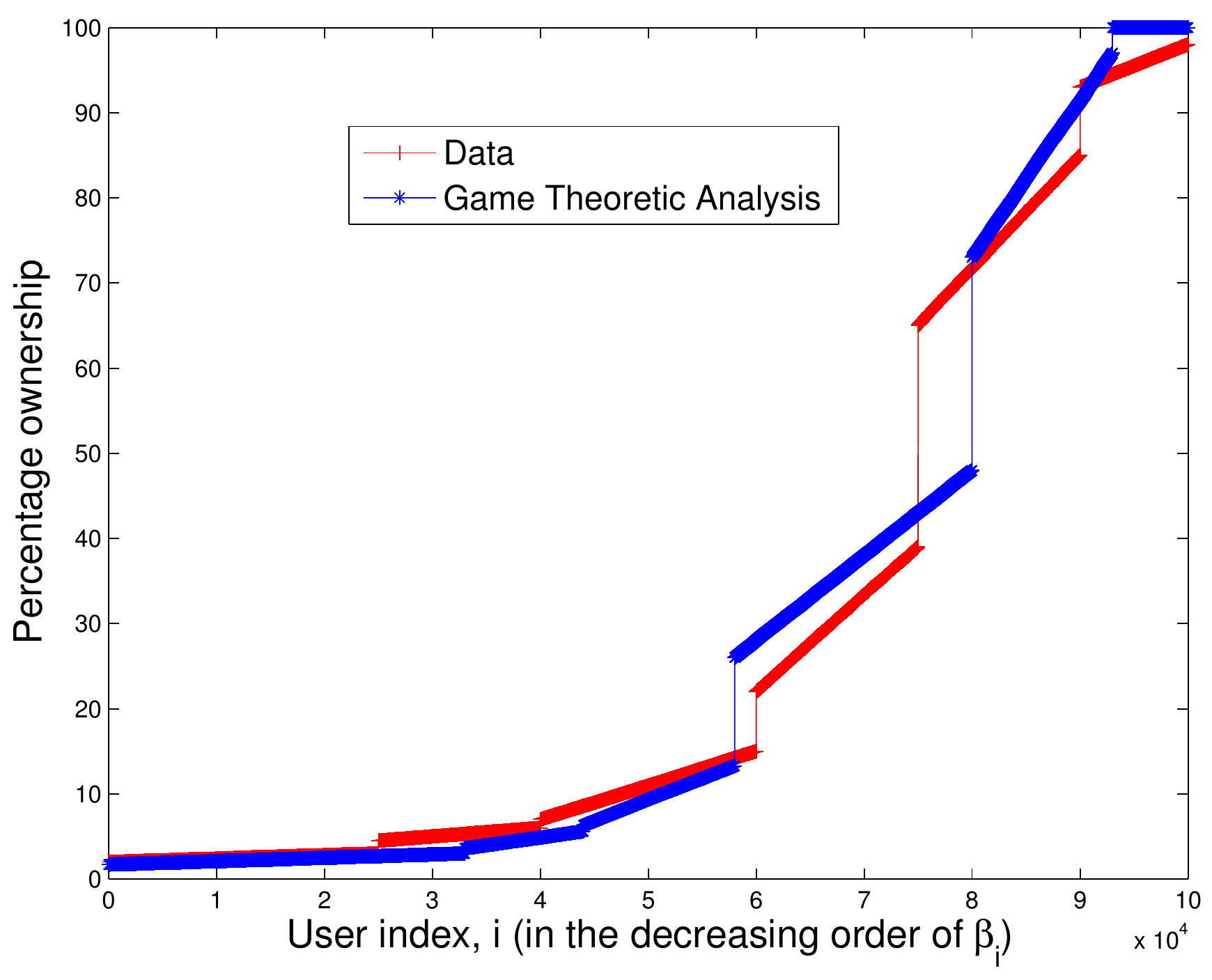}
\label{fig:comparesecond}}
\caption{\label{fig:comparedata} The percentage ownership obtained from 
the data and that obtained from Equation 
(\ref{eqn:bistar}) for 
for 864 multi-contributor articles in our sample.
Results represent the aggregated solution for the 864 games (each game solved with the appropriate number 
of editors) and are shown for articles after the end of 
the entire period of study (Fig. \ref{fig:compare}),
the year of inception or the first year
(Fig. \ref{fig:comparefirst}) and the second year (Fig. \ref{fig:comparesecond}).
Contributors are indexed according to the 
decreasing order of edit size, $\beta_i$'s.}
\end{center}
\end{figure*}

Our findings show that the model's prediction regarding contributors' 
fractional ownership are almost identical to the empirical findings. Please see 
Figure \ref{fig:compare} for the results for the entire study's period. 
Looking at the temporal brackets, we observe that already at the end of 
the year of inception (Year 1), the process converges and our model's 
predictions closely match the empirical data (see Figure \ref{fig:comparefirst}). 
Figure \ref{fig:comparesecond} shows that from Year 2 onward, the results are 
almost identical to the findings for the entire period. 
These results suggest that as articles mature and the co-production process 
stabilizes, our empirical results become more aligned with the model's prediction. 
In order to further test for the effect of articles' maturity, we analyzed the alignment 
between model's prediction and empirics by the number of contributors per article. 
We found that the alignment increases as the number of contributors per article 
grows (stabilizing at roughly 30 contributors per article).
We measured the Pearson's correlation coefficient value between the 
fractional ownership obtained by the game theoretic analysis and that obtained 
from the measured data ($2$ years from inception, and onward), and found it to be 
$0.873$, thus demonstrating, that the data closely follows our model's predictions.
The stability of our findings across various stages of an article's 
life cycle solidifies the validity of our model, demonstrating that despite 
simplifying assumptions and use of a static model, we are 
able to capture the temporal dynamic underlying competition of an article's contents.
In order to verify the robustness of our model, we have re-ran our analysis, this 
time when excluding non-human editors (i.e. ``bots''). Our findings indicate that the 
pattern of results - for both the entire-period and temporal bracketing analyses - 
remains the same.

Trying to get a better sense of the (small) 
discrepancies in ownership values between the model's prediction 
and empirical data, we explored whether the differences could be 
offset by establishing a 
linear fit function mapping. Let 
$\mathbf{a}\define
\left [ 
\begin{array}{ccccc}
a_1 & a_2 & a_3 & \cdots & a_N
\end{array}
\right ]$ (representing the ownership of the 
contributors obtained by the game theoretic 
analysis) and $\mathbf{d}\define
\left [ 
\begin{array}{ccccc}
d_1 & d_2 & d_3 & \cdots & d_N
\end{array}
\right ]$ (representing the ownership of the 
contributors obtained from the empirical data). 
For each Wikipedia article, we fit a function 
\BEQA\label{eqn:linfit}
\begin{array}{cc}
\hat{d}_i=\rho a_i+\delta, & 1\leq i\leq N,
\end{array}
\EEQA
where the parameters, $\rho$ and $\delta$, are obtained
by the method of least squares \citep{Meyer2000}. 
We randomly selected 300 articles as training data in 
supervised learning to determine the linear regression 
coefficients, repeating the training/testing split 1000 times. 
The average error for the training data was 8\%. We then 
ran a linear regression on the remaining 564 pages and tested 
the linear fit as well as its significance by measuring the
$p-$value. The error was found to be between 
11-16\% and the linear regression was statistically significant 
at $p < 0.05$. 
The implication of this result   is that the game theoretic analysis 
models the contributors' 
interactions in Wikipedia accurately up to a linear scaling factor.

\section{Discussion and Conclusion}\label{sec:concl}
In this paper we developed a game-theoretic model of competition and governance in 
Wikipedia and corroborated the model through an empirical study. 
We stress that we did not intend our game-theoretic model to be comprehensive; 
socio-technical systems such as Wikipedia are much too complex to be fully captured 
through a game-theoretic model. Instead, we attempted to model few essential features 
of Wikipedia's co-production process, namely contributors' competition over content 
ownership and the community's strive to ensure a balanced position.
The results of our model, supported by findings from an analysis 
of $219,811$ distinct contributors co-producing $864$ representative articles 
over an eleven-year period, seem counter-intuitive. Namely, our findings 
indicate that under the conditions of the community's strive to ensure neutrality, 
the ``creators'' who on average make large contributions (and thus exert high 
effort per editing activity) end up owning relatively little of the article's contents. 
Furthermore, asymptotically, only contributions made by ``curators'', whose edit
sizes (and effort) are below that expended on average by the group of contributors 
working on the article, survive the on-going changes to the article's content.

How could this result be explained? 
One line of explanation for this points to 
community's effort to ensure neutrality \citep{GreensteinZhu,GreensteinZhu2017,Young2020}. 
The ``creators'' making large contributions and taking 
ownership of large portions of the 
article may potentially pose a threat to the article's neutrality. In response, the community 
works collectively to fight such manipulations  and restores neutrality. In 
the long run, the result is a more neutral content 
\citep{Stvilia2008,GreensteinZhu2012}.
Not withstanding the role of 
community efforts to ensure neutrality, our results may also be explained 
by paying attention to the profile of those making the large contributions. 
Prior studies have identified the ``creators'' 
\citep{Kim2017,Rainie2012,Roque2016,Hull2013,Hill2017}
or ``content-oriented'' contributors as those 
mainly interested in explicating their knowledge of the topic, and much 
less in Wikipedia's internal processes. In contrast, the ``curators'' 
(also referred to as ``community-oriented'' contributors \citep{Arazy2011}), 
who engage in refining, shaping, and refactoring of 
content \citep{Butler2007} and act as 
``janitors of knowledge'' \citep{NiedererVanDijck2010}  
own larger portions of articles. 
Both these lines of explanation highlight the effectiveness 
of Wikipedia's governance. 

A second powerful result of this study is in demonstrating that 
that the community's efforts to govern content creation and ensure neutrality,
although essential for maintaining a balanced position, 
should be curtailed. The reason is that 
when the ``tax'' imposed on contributors in terms of complying with NPOV
norms, policies and procedures is too high it outweighs the benefits 
associated with content ownership, such that contributors stop 
competing for ownership (and in effect, co-production is stalled). 
Prior studies have documented the growing bureaucracy of Wikipedia 
\citep{Butler2008}, and have suggested that the increasing 
complexity of Wikipedia's governance structure is deterring 
newcomers and may eventually lead to the decline of Wikipedia 
\citep{Halfaker2012}. Our results provide an analytical proof for 
the risk associated with increased governance cost and highlight the 
need to balance the cost and benefits associated with the governance 
of online production communities. 

In sum, our study makes important contributions to research on peer production, 
First we shed light on the
competitive dynamics underlying peer production.
Second, we show how Wikipedia's governance and in particular, the 
community's efforts to maintain neutrality, affect contributors' behavior 
and consequently, the quality of the co-produced articles. 
Although there have been studies using game theory and network
ties to study collaboration in social networks in general \citep{Hanaki2007} and
applied to Wikipedia in particular \citep{Zhang2012}, 
we are not aware of prior research in this area that linked 
governance to individuals' actions. 

Notwithstanding these contributions, our work could be extended
in several directions, for example by employing alternative measures 
for our model's constructs (e.g. alternative measures of articles' 
objectivity, for example through an analysis of Wikipedia's discussion 
pages) and extending the model to include additional constructs 
related to contributors (for example: contributors' intent and their 
psychological ownership, their compliance with Wikipedia's policies, 
or the phase in an article's life when they concentrate their edits) and 
related to the articles (the extent to which articles are controversial or 
the importance of the knowledge-based product). Future research 
could also attempt to empirically validate our model's prediction that 
excessive levels of governance impede peer-production.

An additional important contribution of this study is in applying game theory to 
investigate peer production in Wikipedia. 
Prior studies have called for the development of new economic models 
\citep{LernerTirole2002} and in particular, the application of game 
theory to investigate emerging socio-technical systems such as Wikipedia 
\citep{Conte2012}.
Few studies have used game-theoretic models to compare the efficiency of closed-
and open-source software development regimes
\citep{BaldwinClark2006}, 
investigate 
the competitive dynamics of the process by which participants provide 
feedback on each other's work (i.e. rating and reputation systems) 
\citep{Dellarocas2003}, and to study co-production in user
generated content \citep{GhoshMcAfee2011},
crowd sourcing  \citep{HortonChilton2010} and Wikipedia \citep{Anand2013}. 
Our study extends prior work in the area by 
using game theory to shed light on the role governance mechanisms play in 
moderating the competition between contributors to peer production. 
Our game-theoretic model provides insights into the complexities of 
cooperation and competition in peer-production, making several 
simplifying assumptions. 

Whereas our empirical analysis corroborated 
the game-theoretic model, we did observe small discrepancies between 
the model's predictions and the empirical data. One
possible explanation for these differences is that our model provided only an 
approximation of Wikipedia's complexities. Future research could help make 
progress: such work could relax some of our model's assumptions, for example 
model learning effects, capture other important qualities of articles such as 
accuracy or completeness, explore multi-stage games, and consider various 
classes of contributors that differ in their roles and goals. 

In particular, we foresee three avenues for extending our study. First, in our 
model a contributor owns his newly-added content thus impacting others' relative 
ownership. A more sophisticated model could also account for overwriting 
(or deleting) another contributor's content, thus directly impacting the ownership 
of a specific other. A challenge for such a model would be to handle the complexity 
associated with numerous pair-wise relationships. 
Second, our model could possibly be extended to capture a tighter linkage between 
the leader-follower game (community governance) and the non-cooperative game 
(competition over content ownership), for example by linking the effectiveness of the 
community's governance to the aggregate of contributors' governance effort.
Third, future work could represent cooperative dynamics within Wikipedia. Although 
our model indirectly accounts for cooperation in that each player in the game could 
be seen to represent a group of collaborative similar-minded contributors, a more 
complete model could directly model cooperative behavior.

In conclusion, we believe that game theory can 
reveal deep insights into the complex dynamics underlying peer production. 
While game theory, and in particular leader-follower game,
has already been employed in the context of allocation of divisible resources 
\citep{Basar2006}, the application of such techniques to community-based 
peer production is novel.
Future research directions for game theory in this area include: 
the use of cooperative game 
theory \citep{Branzei2008} to study the stability of the production process; 
applying coalitional game models \citep{RayVohra1999} to 
analyze how competing coalitions of contributors emerge, 
applying resilience and immunity models \citep{Halpern2008} to investigate how 
the peer production process performs in cases of deviations from expected 
behavior; 
using evolutionary game theory \citep{Weibull1997} 
to study how unexpected behaviors emerge as a result of sequential 
actions by different contributors; to develop models that would account 
for the relationship between individual's governance work and the community's 
overall governance level and using algorithm and mechanism design 
\citep{Jackson2000} 
to determine the most effective community governance structures. 
\bibliographystyle{ormsv080}

\clearpage
\appendices
\section{Notations Used in the Analysis of this Research}\label{app:notations}

\begin{table}[htbp]
\caption{\label{tab:notations} List of notations used in the analysis in this research} 
\begin{center}
\begin{tabular}{|c|l|}
\hline
Notation & Description\\
\hline
$N$ & Number of contributors for a focal Wikipedia article\\
\hline
$x_i$ & Amount of contribution by the $i^{th}$ contributor to the focal article\\
\hline
$c_i$ & Fractional (content) ownership of the $i^{th}$ contributor in the focal article\\
\hline
$f_i$ & Fixed cost incurred by the $i^{th}$ contributor\\
& (by participation in peer production and governance work)\\
\hline
$t$ & Level of neutrality enforcement\\
\hline
$\beta_i$ & Average size of the $i^{th}$ contributor's edits \\
& (across all articles he contributed to)\\
\hline
$E(\mbox{\boldmath{$\beta$}})=\frac{1}{N}\sum_{i=1}^N\beta_i$ & Average size of edits for all those contributing to the focal article\\
\hline
$u_i=c_i-(L\beta_i+t)x_i-f_i$ & Utility of the $i^{th}$ contributor in the focal article\\
\hline
$x_i^*$ &  Optimal amount of contribution by the $i^{th}$ contributor\\
& in the focal article which maximizes $u_i$ with respect to $x_i$\\
\hline
$c_i^*$ & Fractional content ownership of the $i^{th}$ contributor within the\\
& focal article when she makes the optimal contribution, $x_i^*$\\
\hline
$\alpha_i$ & A simplified notation to represent $\frac{1}{L\beta_i+t}$\\
\hline
$\mathbf{D}$ & An $N\times N$ diagonal matrix with $N$, $0$, $\cdots$, $0$ along the diagonal\\
\hline
$\mathbf{D}_{\alpha}$ & A diagonal matrix with $\alpha_1$, $\alpha_2$, $\cdots$, $\alpha_N$ along the diagonal\\
\hline
$\mathbf{1}$ & A column vector in which all entries are $1$, whose transpose is $\mathbf{1}^T$\\
\hline
$\mathbf{P}$ & A matrix whose columns are orthonormal eigen vectors of $\mathbf{1}\mathbf{1}^T$\\
\hline
$H(t)$ & Entropy of an article with governance level, $t$\\
\hline
$s_i$ & Total number of sentences owned by the $i^{th}$ contributor in the focal article\\
\hline
$\zeta_i$ & Total number of edits made by the $i^{th}$ contributor \\
&over all articles in our sample\\
\hline
\end{tabular}
\end{center}
\end{table}

\clearpage
\section{Nash Equilibrium of the Non-Cooperative Game Between the Contributors}\label{app:NashEq}
In this section, we detail the steps in obtaining the optimal contributions Re-writing Equation (\ref{eqn:firstderivative}), 
\begin{eqnarray}\label{eqn:omegai}
\begin{array}{cc}
\left (\sum_{j=1}^N x_j^*\right )^2
-\alpha_i\sum_{j=1\atop j\neq i}^Nx_j^*=0,
& \forall N,
\end{array}
\end{eqnarray}
where $\alpha_i\define \frac{1}{t+L\beta_i}$.
Equation (\ref{eqn:omegai}) can be written as
\begin{eqnarray} \label{eqn:modifiedmatrixform}
\left (\mathbf{x}^*\right )^T
\mathbf{1}\mathbf{1}^T
\mathbf{x}^*\mathbf{1}-
\mathbf{D}_{\alpha}\left (\mathbf{1}\mathbf{1}^T-\mathbf{I}\right )
\mathbf{x}^*=\mathbf{0},
\end{eqnarray}
where $(.)^T$ represents the transpose of
a vector or a matrix, $\mathbf{D}_\alpha$ is the diagonal
matrix $\mathbf{diag}\left (\alpha_1,
\alpha_2,\cdots,\alpha_N\right )$,
$\mathbf{1}$ is the column vector in
which all entries are one, $\mathbf{0}$ is the column
vector in which all entries are zero and $\mathbf{I}$
is the identity matrix. 

It can be easily verified the vectors,
$\mathbf{y}_1=\frac{1}{\sqrt{N}}\mathbf{1}$
and for $j=2$, $3$, $\cdots$, $N$,
$\mathbf{y}_j=\left [y_{kj}\right ]_{1\leq k\leq N}$,
where
\begin{eqnarray}\label{eqn:yj}
y_{kj}=\left \{
\begin{array}{cc}
-\frac{1}{\sqrt{j(j-1)}} & k<j\\
\frac{j-1}{\sqrt{j(j-1)}} & k=j\\
0 & k>j,
\end{array}
\right .
\end{eqnarray}
form a set of orthonormal eigen vectors
to the matrix, $\mathbf{1}\mathbf{1}^T$.
The eigen value corresponding to $\mathbf{y}_1$
is $N$ and those corresponding to $\mathbf{y}_2,\cdots,\mathbf{y}_N$
are $0$s.
Let 
\begin{eqnarray}\label{eqn:defnP}
\mathbf{P}=\left [
\mathbf{y}_1\vert\mathbf{y}_2\vert\cdots\vert\mathbf{y}_N\right ].
\end{eqnarray}
Then, $\mathbf{P}$ is an orthogonal matrix and by orthogonality transformation 
\citep{Meyer2000}, 
\begin{eqnarray}\label{eqn:orthogonality}
\mathbf{P}^T\mathbf{1}\mathbf{1}^T\mathbf{P}=\mathbf{D}=
\mbox{diag}\left (N, 0, 0, \cdots, 0\right ).
\end{eqnarray}
Let $\mathbf{z}=\left [
\begin{array}{cccccc}
z_1 & z_2 & z_3 & \cdots & z_{N-1} & z_N
\end{array}
\right ]^T$.
Since the eigen vectors of a matrix form a basis for the
$N-$dimensional sub-space \citep{Meyer2000},
the vector, $\mathbf{x}^*$, can be written as
$\mathbf{x}^*=\mathbf{P}\mathbf{z}$. In other words, from Equations (\ref{eqn:yj}) and (\ref{eqn:defnP}),
\begin{eqnarray}\label{eqn:x1fromz}
x_1^*=\frac{z_1}{\sqrt{N}}-\sum_{j=2}^N\frac{z_j}{\sqrt{j(j-1)}},\\
x_k^*=\frac{z_1}{\sqrt{N}}+\frac{(k-1) z_k}{\sqrt{k(k-1)}}-\sum_{j=k+1}^N\frac{x_j}{\sqrt{j(j-1)}},\mbox{ }2\leq k\leq N-1,\label{eqn:xkfromz}\\
x_N^*=\frac{z_1}{\sqrt{N}}+\frac{(N-1)z_N}{\sqrt{N(N-1)}}\label{xNfromz}.
\end{eqnarray}
Intuitively, the vector, $\mathbf{z}$ is a linear transformation of the set of variables in the vector, $\mathbf{x}^*$,
which enable solving the set of Equations characterized by (\ref{eqn:modifiedmatrixform}), using an 
approach similar to that outlined in \citep{Anand2014}, described by the following steps.
\begin{itemize}
\item Using  $\mathbf{x}^*=\mathbf{P}\mathbf{z}$ in Eqn. (\ref{eqn:modifiedmatrixform})
and Equation (\ref{eqn:orthogonality}), we obtain
\begin{eqnarray}\label{eqn:nonlineZ}
\mathbf{z}^T\mathbf{D}\mathbf{z}\mathbf{1}
-\mathbf{D}_\alpha\left (\mathbf{1}\mathbf{1}^T-\mathbf{I}\right)
\mathbf{P}\mathbf{z}=\mathbf{0}.
\end{eqnarray}
 \item The above is a set of non-linear equations in
$\mathbf{z}$, in which the $k^{th}$ equation depends
on $z_1$ and $z_j$, $k\leq j\leq N$. Solving the
non-linear equations by backward substitution \citep{Meyer2000}, 
$z_k$, $2\leq k\leq N$ can be written in
terms of $z_1$ as 
\begin{eqnarray}\label{eqn:xkx1}
\frac{z_k}{\sqrt{k(k-1)}}=
\frac{Nz_1^2}{k(k-1)}\left [
\frac{k}{\alpha_k}+\sum_{j=k+1}^N\frac{1}{\alpha_j}\right ]
-\frac{z_1}{\sqrt{N}}\frac{N(N-1)}{k(k-1)}.
\end{eqnarray}
\item Using Equation (\ref{eqn:xkx1}) to replace all $z_k$'s 
in terms of $z_1$ in the set of non-linear equations 
in Eqn. (\ref{eqn:nonlineZ}), $z_1$ can be obtained as
\begin{eqnarray} \label{eqn:expz1}
z_1=\frac{N-1}{\sqrt{N}}
\frac{1}{G}, 
\end{eqnarray}
where
\begin{eqnarray}\label{eqn:defnG}
G\define\sum_{j=1}^N\frac{1}{\alpha_j}.
\end{eqnarray}
\item Combining Equations (\ref{eqn:xkx1}) and (\ref{eqn:expz1}),
for $2\leq k\leq N$,
\begin{eqnarray}\label{eqn:finalexpressionxk}
\frac{z_k}{\sqrt{k(k-1)}}=
\frac{(N-1)^2}{k(k-1)}G^{-1}
\left [G^{-1}\left (\frac{k}{\alpha_k}+
\sum_{j=k+1}^N\frac{1}{\alpha_j}\right )-1\right ] 
\end{eqnarray}
\item  Using Equations (\ref{eqn:x1fromz})-(\ref{xNfromz})
and $\alpha_i=\frac{1}{t+L\beta_i}$ in
Equation (\ref{eqn:expz1}) and Equation (\ref{eqn:finalexpressionxk}), 
the unique Nash equilibrium, $\mathbf{x}^*$ can be obtained as
given by Equation (\ref{eqn:bistar}).
\end{itemize}
\clearpage
\section{Validation of the Objective Function for Neutrality Enforcement}\label{app:JMISDataSet}
We sought to validate reasoning underlying the objective function. 
Namely, we sought evidence that the goal of the Wikipedia community is 
indeed to maximize neutrality (i.e. entropy) in content ownership. 
Optimally, we would have been able to compare the entropy in content ownership 
to article's objectivity. However, such a metric of objectivity is not available for our 
dataset, especially when we consider that our empirical validation requires this 
metric for each article revision. Instead, we employed a more general metric of 
content quality, recognizing the limitation that objectivity is but only one dimension 
of content quality. We, thus, compared the entropy in content ownership for all 
articles in our sample with article?s quality at the cut-off date of our study. 
Our entropy measure is normalized (i.e. each article's entropy is calculated in 
relation to the maximum entropy of any article, which is 1), so as to control 
in variations in the number of contributors per article.

\begin{figure*}[htbp]
\centering
\includegraphics[width=3.75in]{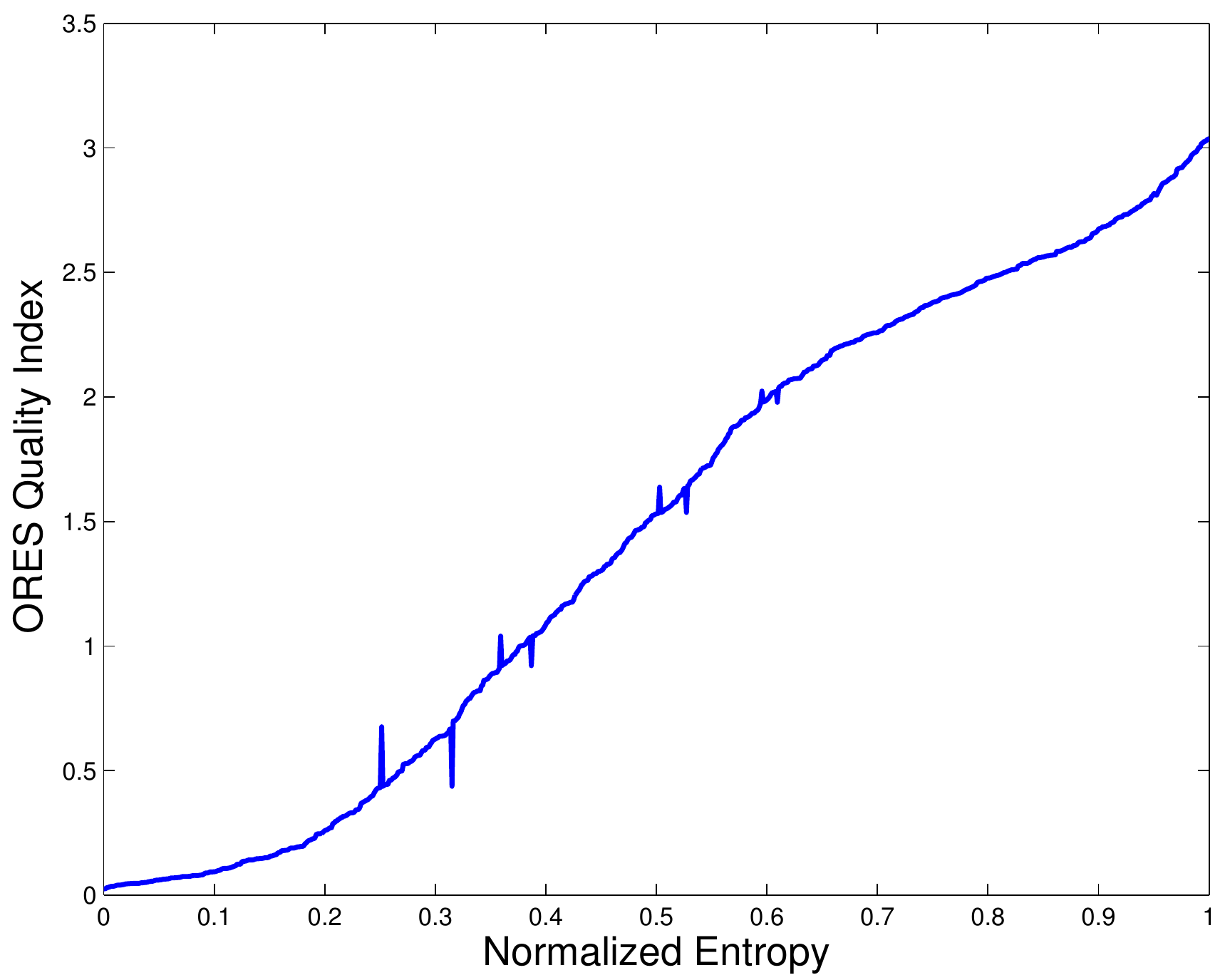}
\caption{\label{fig:overallquality864articles} The quality index for the 864 articles analyzed in this study as a function 
of the entropy in articles' content ownership.There is a strong positive
correlation (Pearson correlation co-efficient $\approx 0.82$) 
between the entropy of an article and its quality. }
\end{figure*}

Deriving such a content quality metric is not an easy task. Although the Wikipedia 
community rank article's quality, content on articles in a continuous flux, and the 
community is not able to rank re-visit the quality of each article revision. 
Furthermore, only a small portion of the 
articles in our sample have been ranked by the Wikipedia community. 
Consequently, 
we chose an alternative method for estimating articles' quality, building on the recent 
work of the Wikipedia Foundation to develop an automated tool for estimating articles' quality. 
The ORES (Objective Revision Evaluation Service) service is intended to be scalable, easily 
extensible and responsive \citep{DangIgnat2016,Sarabadani2017}. ORES uses 
automatic machine-learning methods.
ORES is trained on Wikipedia's community-based article
scoring procedure, which stresses the goals of content reliability, verifiability, 
objectivity, and importance, and assigns articles with quality grades (from lowest to 
highest: Stub, Start, C grade, B grade, Good Article (GA), and Featured Article (FA)) . 
We used the ORES API to extract quality estimates for each Wikipedia article in our 
sample of 864 articles at the date of our study's cut-off. The ORES API outputs 
probabilities for having the article associated with each of the quality grades. 
Finally, the probabilities were weighted using the scheme of [Stub = 0, Start = 1, C = 2, B = 3, GA = 4, FA = 5] 
to produce a score between 0 and 5. 
As shown in Figure \ref{fig:overallquality864articles}, we found a strong correlation 
(Pearson correlation coefficient $\approx 0.82$) 
between articles' quality and the entropy in content ownership, 
suggesting that maximizing entropy is a reasonable objective function to maximize.

In order to provide an additional verification for our objective function 
(i.e. maximizing entropy for determining the optimal level of governance, $t$), 
we sought additional evidence that the goal of the Wikipedia community is indeed to 
maximize neutrality (i.e. entropy)  in content ownership. We employed a set of $96$ 
Wikipedia articles used in \citep{Arazy2011}\footnote{This data set with articles' quality 
metrics was obtained through direct correspondence with the authors.}, which sampled 
articles from Wikipedia's various topical categories. On average an article in this set 
was edited by $49$ contributors and went through $91$ revisions. This data set provides 
a robust metric of objectivity, along with additional metrics of information quality along the 
dimensions of: accuracy, completeness, and representation (on a $7-$point Likert scale). 
These metrics were determined through a three-phased approach: $(1)$ manual analysis by student 
assessors as part of a course assignment (multiple assessors per article); $(2)$ 
independent manual analysis by senior university librarians comparing the articles 
to external sources (three assessors per article); and $(3)$ consensus between the 
three librarians using a Delphi approach. The outcome variable for this data set was 
the average of consensus score for the various quality metrics, on a $1-7$ scale. Using 
the same method applied to our study's primary sample, 
we calculated the ownership of each article and the entropy in contributors' content ownership. 
Results from our analysis show a strong correlation between entropy in ownership and article 
quality (Pearson coefficient $\approx 0.78$), further supporting our definition of the objective 
function (for determining optimal governance level, $t$) as the maximization of articles' 
entropy (in terms of contributors' content ownership).
\clearpage
\section{Determining the Optimal value of Neutrality Enforcement, $t$}\label{app:Optimalt}
We now present the detailed analysis to show the fact that the optimal value
of $t$ that maximizes the entropy of a page is $t\rightarrow\infty$.
From Equation
\BEQA\label{eqn:intermediate}
\frac{\partial H}{\partial z}=-\sum_{i=1}^N\frac{\partial c_i^*}{\partial z}(1+\mbox{ln }c_i^*). 
\EEQA
Therefore, the condition $\frac{\partial H}{\partial z}=0$ yields the condition,
\BEQA\label{eqn:condition}
\sum_{i=1}^N \frac{\partial c_i^*}{\partial t} \mbox{ln }c_i^*=-\sum_{i=1}^N \frac{\partial c_i^*}{\partial t}\mbox{ln }c_i^*.
\EEQA
From Equation (\ref{eqn:uistar}),
\BEQA\label{eqn:firstderivativeci}
\frac{\partial c_i^*}{\partial z}=\frac{(N-1)\left ( NL\beta_i-\sum_{j=1}^N L\beta_j\right )}{\left (Nz+\sum_{j=1}^N L\beta_j\right )^2} 
=\frac{N(N-1)\left (L\beta_i-LE[\mbox{\boldmath{$\beta$}}]\right )}{\left (Nz+\sum_{j=1}^N L\beta_j\right )^2},
\EEQA 
\BEQA\label{eqn:sumderivative}
\mbox{i.e., } \frac{\partial H}{\partial z}=0\Rightarrow \sum_{i=1}^N \frac{\partial c_i^*}{\partial z}=0.
\EEQA
Putting Equation (\ref{eqn:sumderivative}) in Equation (\ref{eqn:condition}) and simplifying,
the optimal neutrality enforcement, $t$ is obtained as the value of $z$ that solves  
\BEQA\label{eqn:finalexpressiont}
\sum_{i=1}^N[(\beta_i-E(\mbox{\boldmath{$\beta$}})]\mbox{ln } c_i^*(z) =0.
\EEQA
From the observations listed in Section \ref{subsec:ncgame}, if $\beta_i>E(\mbox{\boldmath{$\beta$}})$, then
$c_i^*=0$, i.e., $ln c_i^*\rightarrow-\infty$. $0<c_i^*<1$ when $\beta_i<E(\mbox{\boldmath{$\beta$}})$, i.e.,
ln $c_i^*<0$. Therefore, the value of $t$ that satisfies Equation (\ref{eqn:finalexpressiont}) is the value
that makes ln $c_i^*$ go to $-\infty$, i.e., make $c_i^*\rightarrow 0$, i.e., $t=\infty$, from Equation (\ref{eqn:uistar}).  
Therefore, the value of $t$ that maximizes $H(t)$ is $t\rightarrow\infty$.

\end{document}